\documentclass{evn2004}
\usepackage{txfonts,graphicx}
\usepackage{natbib}

\def\mmdo{{M$_{\rm MDO}$}}

\def\H0{{\rm ~km~s^{-1}~Mpc^{-1}}}

\def\ledd{${l_{\rm Edd}}$}

\def\ergsec{{\rm ~erg~s^{-1}}}

\def\hii{\ion{H}{II}}
\def\ha{{H$\alpha$}}

\def\.25{0.25 keV\thinspace}

\def\d19{D$\,\leq\,$19~Mpc}
\def\dgtr19{D$\,>\,$19~Mpc}

\let\lesssim=\la
\let\gtrsim=\ga

\begin{document}
   \title{Radio Luminosity Function, Importance of Jet Power, and
          Radio Properties of Nearby Low-Luminosity Active Galactic Nuclei}

   \titlerunning{Radio Properties of LLAGNs}

   \author{Neil  M. Nagar\inst{1}
          \and
          Heino Falcke\inst{2}
          \and
          Andrew S. Wilson\inst{3}
          }

   \institute{Kapteyn Institute, Landleven 12, 9747 AD Groningen, The Netherlands
         \and
	      ASTRON, P.O. Box 2, 7990 AA Dwingeloo, The Netherlands
         \and
	      Department of Astronomy, University of Maryland,
	                   College Park, MD 20742, U.S.A. \\
	                 Adjunct Astronomer, Space Telescope Science Institute,
	                   3700 San Martin Drive, Baltimore, MD 21218, U.S.A.
             }
   \abstract{
We present the completed results of a high resolution radio imaging survey
of all ($\sim$200) low-luminosity active galactic nuclei (LLAGNs) in
the Palomar Spectroscopic Sample of all ($\sim$470) nearby bright northern galaxies.
The high incidence of pc-scale radio nuclei, with implied brightness
temperatures $\gtrsim\,10^7\,$K, and sub-parsec jets argue for
accreting black holes in $\gtrsim$50\% of all LLAGNs; there is no evidence
against \textit{all} LLAGNs being mini-AGNs.
The detected parsec-scale radio nuclei are preferentially found in massive
ellipticals and in type~1 nuclei (i.e. nuclei with broad \ha\ emission).
The radio luminosity function (RLF) of Palomar Sample LLAGNs extends three
orders of magnitude below, and is continuous with, that of `classical' AGNs.
We find marginal evidence for a low-power turnover in the RLF; nevertheless
LLAGNs are responsible for a significant fraction of accretion in the local
universe.
The accretion energy output in LLAGNs is dominated by the energy in the observed
jets rather than the radiated bolometric luminosity. The Palomar LLAGNs follow
the same scaling between jet power and narrow line region (NLR) luminosity as the
parsec to kilo-parsec jets in powerful radio galaxies.
Low accretion rates ($\leq10^{-2}-10^{-6}$ of the Eddington rate) are implied
in both advection- and jet-type models, with evidence for increasing
`radio-loudness' with decreasing Eddington fraction.
The jets are energetically more significant than supernovae in the LLAGN host 
galaxies, and are potentially able to dump sufficient energy into the innermost
parsecs to significantly slow the accretion inflow.
Detailed results can be found in Nagar et al. (2002a) and Nagar et al. (2004, to 
appear in Astronomy \& Astrophysics).
   }

\maketitle
%

\section{Introduction}
\label{secintro}
The debate on the power source of low-luminosity active galactic 
nuclei (LLAGNs, i.e. low-luminosity Seyferts, LINERs, and ``transition'' 
nuclei) is a continuing one. Their low emission-line luminosities 
can be modeled in terms of photoionization by hot, young stars,
by collisional ionization in shocks, or by aging starbursts.
On the other hand, evidence has been accumulating that at least some 
fraction of LLAGNs share characteristics in common with powerful AGNs.
If LLAGNs are truly mini-AGNs then their
much lower accretion luminosities demands either very low accretion 
rates ($\sim\,10^{-8}$ of the Eddington accretion rate)
or radiative efficiencies (the ratio of radiated energy to accreted mass) 
much lower than the typical value of $\sim\,$10\% 
assumed for powerful AGNs.    

One well-known property of some powerful AGNs is a compact 
(sub-parsec), flat-spectrum nuclear radio source, usually interpreted as
the synchrotron self-absorbed base of the jet which fuels
larger-scale radio emission. 
It has been suggested that scaled-down versions of AGN 
jets can produce flat-spectrum radio nuclei in LLAGNs \citep{falbie99}.
Compact nuclear radio emission with a flat to inverted spectrum
is also expected from the accretion inflow in advection-dominated 
(ADAF) or convection-dominated \citep[CDAF;][]{naret00} 
accretion flows, possible forms of accretion onto a black hole at low 
accretion rates.
Flat-spectrum radio sources can also result through thermal emission
from ionized gas in normal \hii\ regions or through free-free
absorption of non-thermal radio emission, a process which probably
occurs in compact nuclear starbursts \citep{conet91}.
The brightness temperature in such starbursts is limited to
log~[T$_{\rm b}~$(K)]~$\lesssim$ 5.
Thus it is necessary to show that T$_{\rm b}$ exceeds this limit before  
accretion onto a black hole can be claimed as the power source. 

How does one distinguish accretion-powered LLAGNs from LLAGNs powered by hot 
stars or supernova shocks? Traditional methods -- broad \ha\ lines, 
unresolved optical or UV sources, broader polarized \ha\ emission,
compact nuclear X-ray emission -- are often ambiguous and
may be affected by viewing geometry,
obscuration, and the signal-to-noise of the observations.
The last problem is exacerbated by the low IR to X-ray luminosities of LLAGNs
and the need to subtract the starlight. 
The radio regime, however, offers several advantages. Gigahertz 
radiation does not suffer the obscuration that affects the UV to IR. 
Also, at tens of gigahertz the problems of free-free absorption can be 
avoided in most cases. Finally, high resolution, high sensitivity 
radio maps can be routinely made with an investment of less than an hour
per source at the Very Large Array (VLA) and 
the Very Long Baseline Array (VLBA).
At their resolutions of $\sim\,$100~milli-arcsec (mas) and $\sim\,$1~mas, 
respectively, it is easy to pick out the AGN, since any other radio emission
from the galaxy is usually resolved out.

Closely related to these theoretical and observational studies of the 
radiation from LLAGNs are the
increasing number of accurate mass determinations for 
``massive dark objects'' (MDO; presumably black holes) in nearby galactic nuclei.
These mass determinations, coupled with the radiated luminosity from the AGN, 
enable a measure of the Eddington fraction, \ledd.
Three quantities -- black hole mass, \ledd, and presumably
the black hole spin -- can then be used to further generalize the physics of,
and consequently `unify', various types of AGNs from Galactic black hole candidates
to the most massive quasars. Here we argue that accounting for the radio jet is
important when estimating \ledd\ in LLAGNs even though the radiated power in the radio
band is not bolometrically dominant.
Our high resolution radio observations of a large number of nearby LLAGNs 
considerably increase the number of LLAGNs with reliable black hole mass
estimates \textit{and} high resolution radio observations, allowing 
a better test of the relationship between these quantities. 

\section{Sample and Radio Observations}
\label{secsample}

The results are
based on LLAGNs selected from the Palomar spectroscopic survey of all
($\sim\,$470) northern galaxies with B$_{\rm T} <$~12.5~mag \citep{hoet97a,hoet03}. 
Of these, roughly 7 are AGNs, 190 are LLAGNs
\citep[using the operational cutoff of L$_{\rm H\alpha}~\leq$ 10$^{40}$ 
erg s$^{-1}$ to distinguish LLAGNs from AGNs;][]{hoet97a},
206 have \hii\ type nuclear spectra, and 53 are absorption line systems.

Several earlier surveys have observed many of the
nearby galaxies which are now in the Palomar Sample \citep[for references
see][]{naget04}.
Since the publishing of comprehensive optical results on the Palomar Sample,
three groups have conducted large radio surveys of the sample. 
Our group has now completed a 0{\farcs}15 resolution 15~GHz (2~cm) VLA survey
of all LLAGNs except some transition nuclei at \dgtr19 
\citep[a total of 162 nuclei observed;][]{naget00,naget02a,naget04}. 
We have then followed up all previously unobserved strong detections with 
the VLBA \citep{falet00,naget02a,naget04}.
\citet{houlv01} and \citet{ulvho01a} have observed all 
Palomar Seyferts at arcsec resolution
at 6~cm and 20~cm and followed up the strong detections at
multiple frequencies with the VLBA \citep{andet04}.
\citet[][and references therein]{filet04} have completed a 5{\arcsec}-0{\farcs}3 
resolution survey of all transition nuclei in the sample with follow up 
VLBA observations of some of the stronger nuclei.

\begin{figure}[ht]
\resizebox{\columnwidth}{!}{
  \includegraphics[bb=75 240 540 390,clip,width=3.6in]{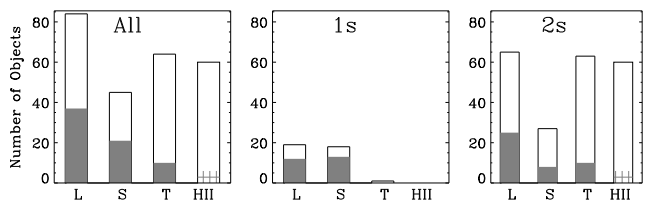}
}
\caption{Detection rate of 15~GHz 150-mas-scale radio nuclei for 
``L''INERs, ``S''eyferts, ``T''ransition, and \hii\ nuclei in the Palomar sample.
Note the higher detection rates of type~1 (i.e. galaxies with broad \ha\ emission)
Seyfert and LINER nuclei.
}
\label{figdetrate}
\end{figure}

\section{Results of the Radio Observations}
\label{secres}

Tables of the complete results of our VLA and VLBA observations of Palomar LLAGNs
appear in \citet{naget04}. 
The detection rate of radio nuclei with the VLA is illustrated in 
Fig.~\ref{figdetrate}. 
The radio luminosities of the detected 2~cm nuclei lie between
10$^{18}$ and 10$^{22}$ W Hz$^{-1}$.
A significant fraction of
the detected 2~cm compact nuclei are in spiral galaxies.
Most of the detected 2~cm nuclear radio sources are compact at the 
0{\farcs}15 resolution (typically 15--25~pc) of our survey: the implied 
brightness temperatures are typically 
T$_b$~$\geq$ 10$^{2.5-4.0}$~K.
The VLBA observations (roughly 43 targets observed)
confirm that all except one (NGC~2655) nuclei with S$_{\rm VLA}^{\rm 15GHz}\,>$ 
2.7~mJy are genuine AGNs with the radio emission coming from mas- or sub-parsec-scales.
About half of the VLBA-detected LLAGNs show one or two sided extensions, reminiscent of
radio `jets.'

\section{Radio Properties}
\label{secderived}

Detailed radio properties of the Palomar LLAGNs appear in \citet{naget02a}
and \citet{naget04}. Here we concentrate on a few key results.

\begin{figure}
\resizebox{3.2in}{!}{
\includegraphics[bb=72 72 302 310,clip]{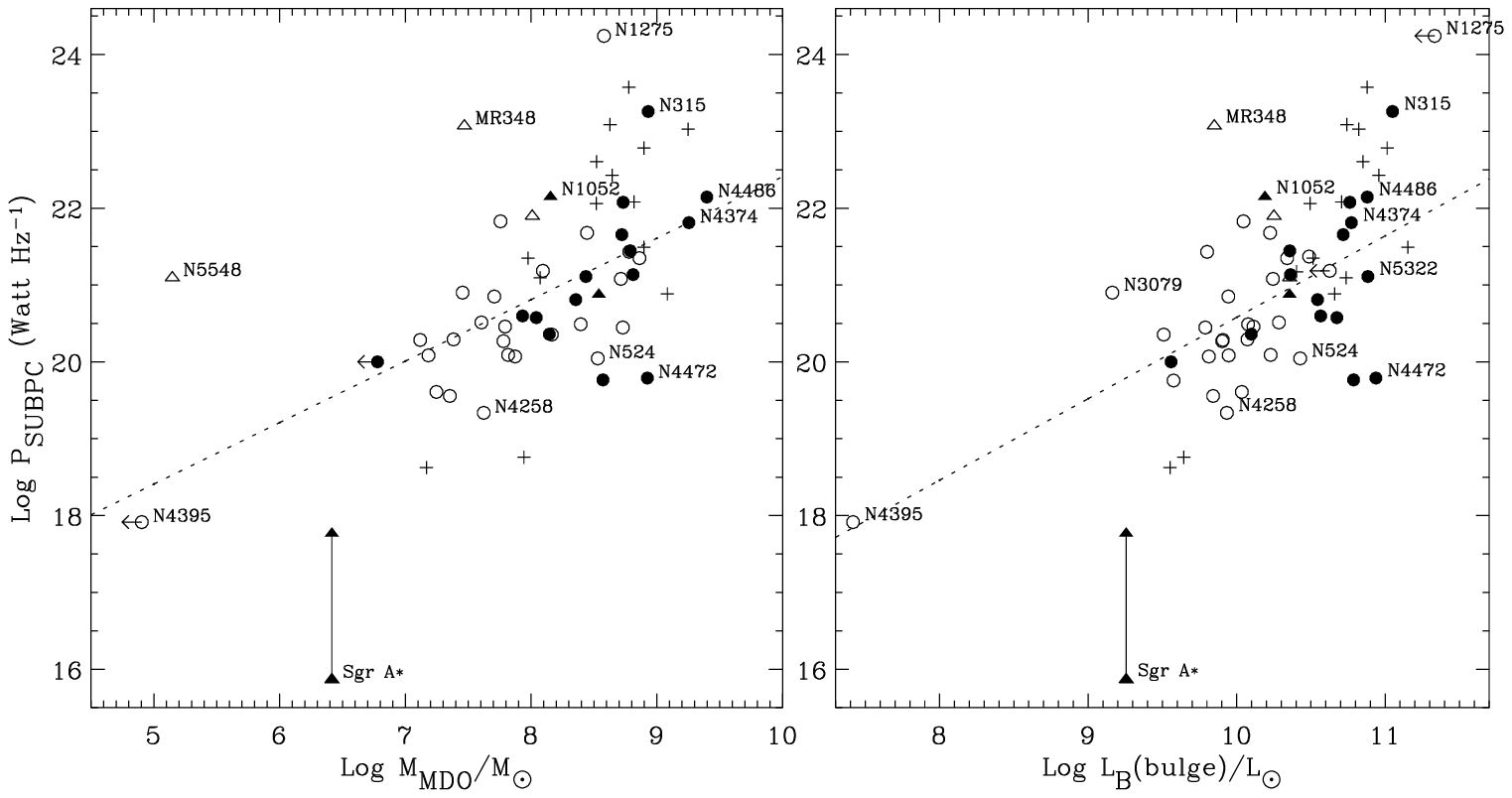}}
\caption{
 A plot of sub-parsec radio power vs. black hole mass.
 Only radio sources relatively unambiguously identified with
 the AGN central engine and with radio fluxes measured
 at resolution $\leq$1~pc ($\leq$5~pc for the crosses) are plotted.
 Palomar LLAGNs and AGNs are plotted as circles and other (LL)AGNs as
 triangles. For these, filled symbols are used for elliptical galaxies.
 Additional nuclei with radio powers measured at resolution between 
 1~pc to 5~pc are shown as crosses.
 The dotted line shows a linear fit to the circles and triangles.
 }
\label{figvlbamdo}
\end{figure}

The radio power is 
correlated with both the black hole mass and the bulge luminosity at the 99.99\% 
significance level. Partial correlation analysis on the two correlations 
yielded the result that each correlation is meaningful even after removing 
the effect of the other correlation \citep{naget02a}.
Here we consider
only nuclei observed with resolution $\leq\,1\,$pc in the radio and for which
one radio component can be relatively unambiguously identified with the 
location of the central engine. This resolution and morphological cutoff 
enables a more accurate measure of the radio emission from only the 
accretion inflow and/or the sub-parsec base of the jet, and helps avoid 
contamination from radio emission originating in knots further out in the jet. 
The latter radio emission is common in LLAGNs and 
often dominates the parsec scale radio emission in Seyferts. In fact 
many Seyferts have several radio sources in the inner parsec, none of which
are clearly identifiable with the central engine.

Fig.~\ref{figvlbamdo} shows the correlation between sub-parsec radio power and
\mmdo.  The plotted circles show the 43 LLAGNs, and the
triangles show 8 additional galaxies which have 
radio nuclei relatively unambiguously identified with
the central engine in maps with resolution better than 1~pc, and
available black hole mass measurements or estimates from $\sigma_c$
(for details see \citet{naget04}).
Linear regression analysis on the circles and triangles in the plot
yields: \\
$\noindent {\rm log\,(P}_{\rm Sub-pc}\,[{\rm W/Hz}])\,=    
     0.8(\pm0.2)\,{\rm log}\,({\rm M}_{\rm MDO}/{\rm M_\odot}) + 14.4   $

\begin{figure*}[ht]
\resizebox{\textwidth}{!}{
   \includegraphics{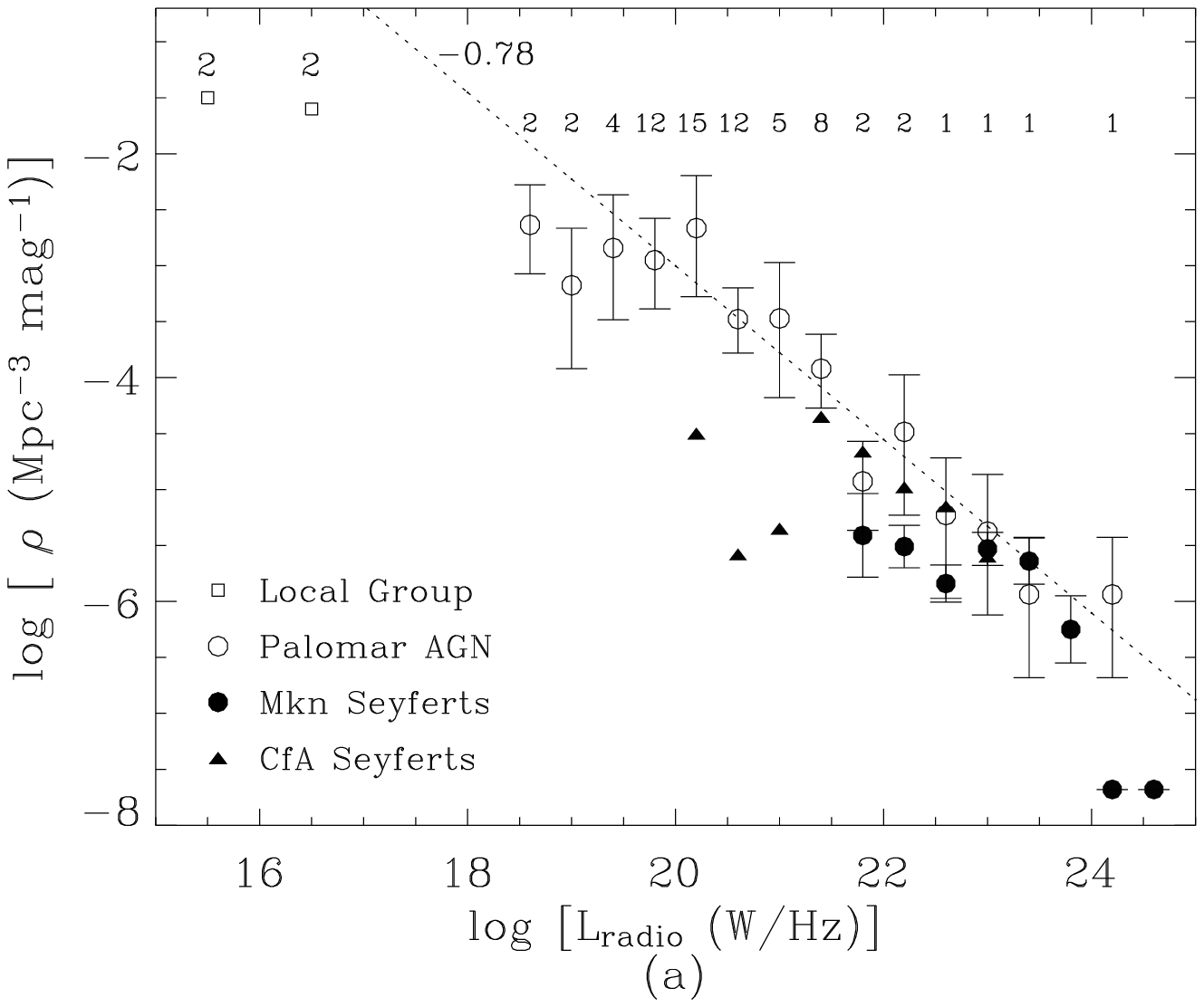}
   \includegraphics{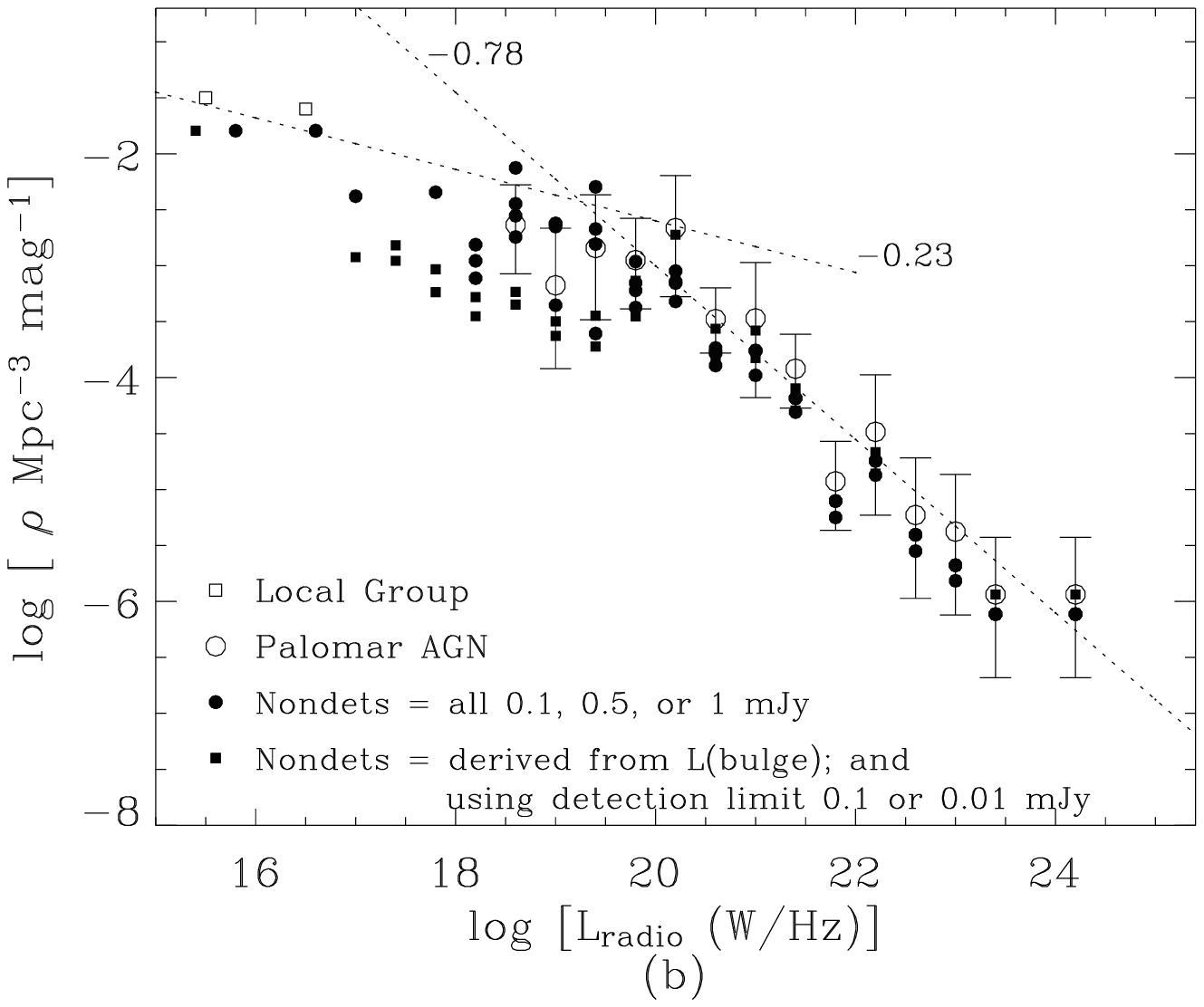}
}
\caption{\textbf{(a)}The 15~GHz radio luminosity function (RLF) of the 150~mas-scale
radio nuclei in the Palomar sample (open circles, with the number of 
galaxies in each bin listed above the symbol). 
For a rough comparison we also
plot the 1.4~GHz RLFs of Markarian Seyferts \citep[1.4~GHz RLF from][]{meuwil84} 
and CfA Seyferts (1.4~GHz RLF calculated by \citet{ulvho01a})
The dashed line is a power-law ($-0.78$) fit to the Palomar nuclear RLF (excluding
the two lowest radio luminosity points).
Also shown is the estimated nuclear RLF of galaxies in the
local group (open squares, with 2 galaxies in each of the two bins);
\textbf{(b)} the open symbols are the same as in the left panel. 
 The filled symbols show the pseudo RLFs for the Palomar sample when 
 some LLAGNs not detected in the radio are taken as radio detections of 1.0, 
 0.5, or 0.1~mJy (and using the same flux as the detection limit; filled circles) or are 
 taken as detections with radio flux derived from the scaling between host galaxy bulge 
 luminosity and radio luminosity and detection limits of 0.1 or 0.01 mJy (filled squares). 
 The dashed line with slope $-$0.23 shows a possible fit (made by eye) 
 to the RLF at the lowest luminosities.
}
\label{figrlf}
\end{figure*}

The nuclear (150~mas-scale) 15~GHz RLF for all 68 radio-detected Palomar 
sample AGNs and LLAGNs is plotted in Fig.~\ref{figrlf}a 
as open circles. 
The RLF has been computed via the bivariate optical-radio luminosity 
function \citep[following the method of][]{meuwil84}.
We emphasize that the nuclear RLF presented here traces
only the very inner AGN jet or accretion inflow, and does not include the
contribution from larger scale radio jets (which are usually not significant 
in LLAGNs).

RLFs at 1.4~GHz and 5~GHz for Palomar Seyferts have been presented in \citet{ulvho01a},
and an RLF (using observations at several frequencies and resolutions) for the complete 
Palomar sample has also been discussed in Filho (2003, $PhD$ thesis).
The RLF we present here 
is in rough agreement with theirs given the errors.
The advantages of the RLF presented here are threefold.
First, it is based on a larger number (68) of radio detections.
Second, it is derived from uniform radio data: all except 13 radio detections
and 21 radio non-detections have their fluxes or upper limits derived from our 
15~GHz (2~cm) VLA A-configuration observations reduced in a uniform way; these
34 exceptions have fluxes or upper limits derived from data of similar 
resolution and frequency.
Third, the radio data were obtained at high resolution and high frequency: 
both these factors reduce the contamination of star-formation-related
emission to the true AGN radio emission, which is specially important at
these low AGN luminosities.

At the highest luminosities the RLF is in good agreement with
that of `classical' Seyferts (Fig.~\ref{figrlf}a). 
We plot the other RLFs without correction for frequency or resolution
(e.g. nuclear vs. total AGN related emission).
This is justified since most of the nuclei in our sample with
multifrequency observations have relatively flat radio spectra from 1.4--15~GHz
\citep{naget01}. Also, the AGN-related radio structures
in these LLAGNs are either sub-arcsec (i.e. the nuclear radio emission is the 
total AGN-related radio emission) or, in a few cases, FR~I-like. Neither case
can be easily compared or corrected to the radio structures in most
Mrk or CfA Seyferts.
At lower luminosities, the sample extends the RLF of powerful AGNs by more than 
three orders of magnitude. A linear (in log-log space) fit to the Palomar nuclear RLF
above 10$^{19}$ Watt Hz$^{-1}$ (i.e. excluding the two lowest luminosity bins; see below) 
yields: \newline
${\log}\,(\rho/{\rm Mpc}^{-3}\,{\rm mag}^{-1}) = (12.5 - 0.78\times\,{\rm log}\,({\rm L}_{\rm radio}\,[{\rm W\,Hz}^{-1}])) $

As we discuss below, a potential fit to the RLF below 10$^{19}$ Watt Hz$^{-1}$
is (with units as above): \newline
${\rm log}\,\rho\, = 2.0\, -\,0.23\,\times\,{\rm log}\,{\rm L}_{\rm radio}$

There is some indication of a low power turnover in the Palomar RLF
(Fig.~\ref{figrlf}a). Admittedly, this apparent turnover is
partly due to the incompleteness of the radio survey, i.e. biased by
the sub-milli-Jansky population which remains undetected.
Nevertheless there are several reasons to believe the presence of such
a turnover, as detailed below and in Fig.~\ref{figrlf}b.
First, and most convincingly, one runs out of bright galaxies:
an extension of the $-0.78$ power law fit to lower luminosities would require
e.g. an LLAGN like Sgr~A* or M~31* to be present in every Mpc$^{-3}$.
To better determine the RLF shape at lower luminosities, we have calculated 
an approximate RLF for the nuclei of the local group of galaxies.
The resulting RLF, plotted with open squares in Fig.~\ref{figrlf}a and b, also
supports a low power break in the Palomar RLF.
As a further test we recalculated the Palomar RLF after converting some or all of 
the radio non-detected LLAGNs into radio detections. All the recomputed pseudo RLFs 
(filled circles and filled squares in Fig.~\ref{figrlf}b)
support a low power break in the Palomar RLF.
The actual shape of the low end of the RLF is uncertain and in Fig.~\ref{figrlf}b
and in the equation above we show a potential power law fit which satisfies the 
current data and extrapolations.

Since LLAGNs lack a `big blue bump', the X-ray has been thought to dominate the bolometric
luminosity \citep{ho99}. 
With typical LLAGNs having hard X-ray luminosities of only $\sim10^{40}$ erg s$^{-1}$ 
or lower, the accretion is highly sub-Eddington \citep{hoet01,terwil03,filet04}.

\begin{figure}[ht]
\resizebox{\columnwidth}{!}{
  \includegraphics{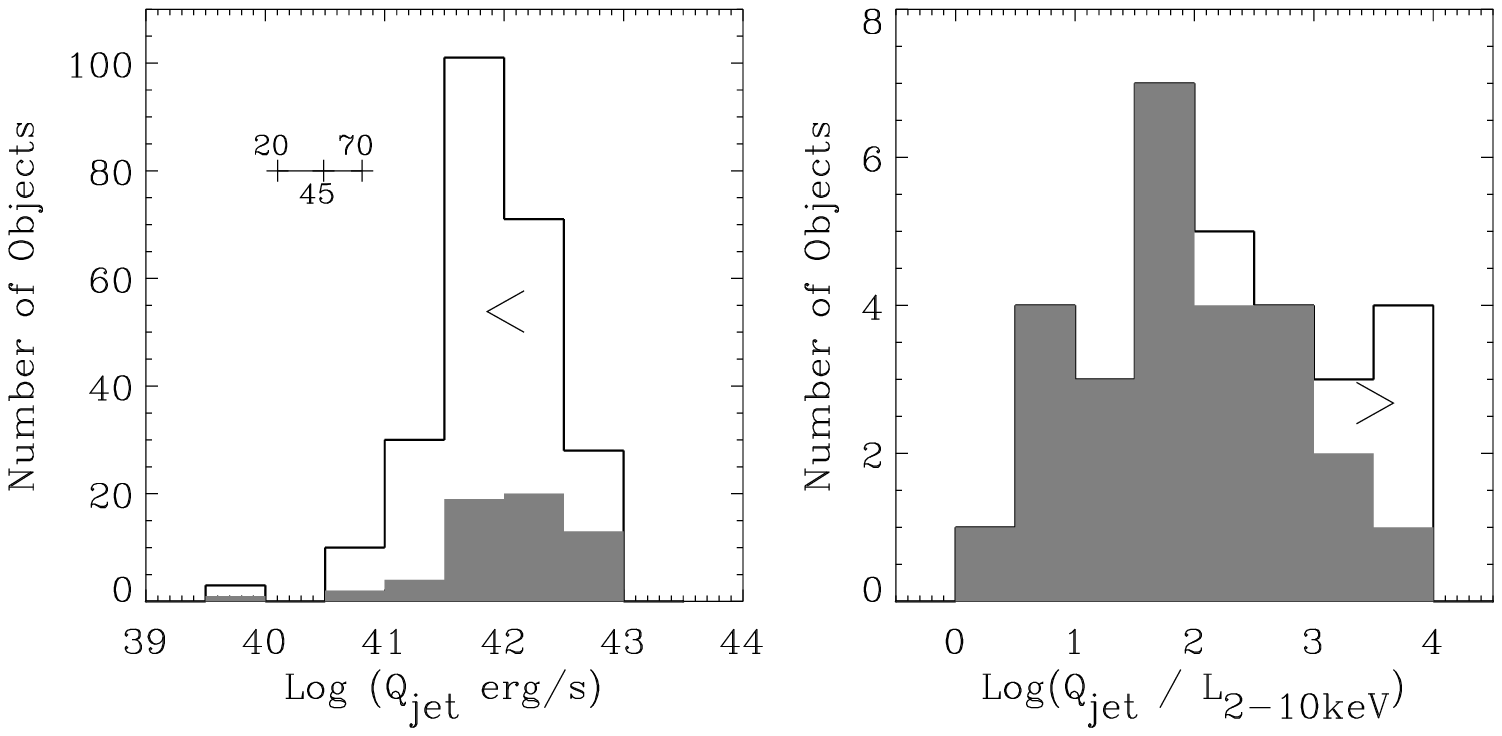}
}
\caption{\textbf{Left:} the implied `jet power' of the radio-detected (gray shaded area) and
radio non-detected (white area) LLAGNs, calculated from Eqn 20 of \citet{falbie99} 
assuming a jet inclination of 45{\degr} to the line of sight. 
The inset illustrates the range of calculated jet powers for three
assumed inclinations: 20{\degr}, 45{\degr}, and 70{\degr}.
\textbf{Right:} log of the ratio of jet power (assuming 45{\degr} inclination)
to X-ray luminosity (in the 2--10~keV band) for
radio detected LLAGNs. The gray and white histograms represent LLAGNs with hard
X-ray detections and upper limits, respectively.
}
\label{figjettoxray}
\end{figure}

Many of the LLAGNs have detected sub-parsec scale 
(and sometimes larger scale) `jets'. 
If the compact radio nuclei and sub-parsec jets represent
emission from the base of a relativistic jet launched close to the black hole,
then the energy in the jet can be quite high.
Equation 20 of \citet{falbie99} - assuming an average inclination
of 45{$\degr$} - predicts jet powers of $10^{41}-10^{43}$ $\ergsec$ 
(Fig.~\ref{figjettoxray}, left panel) for the radio detected LLAGNs. 
For LLAGNs with both hard X-ray and radio luminosity available, this jet power 
greatly exceeds the radiated X-ray luminosity (Fig.~\ref{figjettoxray}, right panel). 
Since L$_{\rm Bol}$ is estimated to be only $\sim$3--15 $\times$ L$_{0.5-10~\rm keV}$ 
for LLAGNs \citep{ho99}, this suggests that the accretion power output is dominated 
by the jet power. 

The energetics of the jet are also important in the context of so called cooling flows 
and in regulating the feedback between galaxy growth and black hole growth. For example,
in most clusters the central CD galaxy has an FR~I radio morphology, with the radio jet 
playing an important role in the above issues and in the global energetics of the cluster.
A comparison of the jet power with the energy injected into the ISM by supernovae
types I and II in the LLAGN host galaxy \citep{naget04} shows that 
the jet power is clearly the major player in the nuclear energetics not only because it
exceeds the total SN kinetic power in almost all cases, but also since its nuclear origin allows
a closer `feedback' to the accretion inflow. 
A significant fraction of the jet energy is expected to be deposited into the central parsecs,
especially in LLAGNs which show pc-scale (usually bent) jets but no larger scale jets 
\citep{naget04}; this can considerably slow down the inner accretion inflow.
Additionally, LLAGNs with kpc-scale jets inject significant energy into the inter-galactic
medium (IGM), and work against any cooling flow.
The most recent of such `feedback' analyses \citep{ostcio04} takes into account the jet
power - though for the more powerful FR~I type jets in CD galaxies. Our results 
show that their models can be extended down to LLAGNs.

\acknowledgements
This work was partly funded by the Dutch research organization
NWO, through a VENI grant to NN, and by NASA through grant NAG513065 to the 
University of Maryland.

\end{document}